\documentclass[11pt]{article}
\usepackage{amsmath,amssymb,amsthm,mathrsfs,booktabs,cite}
\usepackage[margin=1.1in]{geometry}
\usepackage{xcolor}
\usepackage[colorlinks,linkcolor=blue!55!black,citecolor=blue!55!black,urlcolor=blue!55!black]{hyperref}

\newtheorem{theorem}{Theorem}
\newtheorem{proposition}[theorem]{Proposition}
\newtheorem{lemma}[theorem]{Lemma}

\theoremstyle{definition}

\newcommand{\LF}{L_{\mathrm F}}
\newcommand{\LB}{L_{\mathrm B}}
\newcommand{\YF}{Y_{\mathrm F}}
\newcommand{\YB}{Y_{\mathrm B}}
\newcommand{\HH}{\mathcal H}
\newcommand{\PP}{\mathcal P_3^\circ}
\newcommand{\erfc}{\operatorname{erfc}}

\newcommand{\ieee}{\textsc{ieee}-754}

\begin{document}

\title{BCC optimality for the three-dimensional Coulomb lattice
  energy}

\author{C. B. Muratov\footnote{Dipartimento di Matematica,
    Universit\`a di Pisa, Largo Bruno Pontecorvo 5, 56127 Pisa, Italy}
  \and M. Novaga$^*$}

\date{\today}

\maketitle 

\begin{abstract}
  We prove that the body-centered cubic lattice uniquely minimizes the
  renormalized Coulomb energy among three-dimensional Bravais lattices
  of fixed covolume, up to orthogonal transformations. Using Ewald
  summation, we write the energy, to within an additive constant, as a
  sum of two positive, rapidly convergent series in real and
  reciprocal space. Their positivity provides finite lower bounds for
  all competitors. A passage to the dual lattice reduces the problem
  to a face-centered cubic comparison. A short basis vector estimate,
  a certified local Hessian, and interval bounds on a compact
  five-dimensional domain complete the computer-assisted proof.
\end{abstract}

\section{Introduction}
\label{sec:introduction}

In 1934, in his quantum mechanical treatment of an interacting
electron gas in a uniform neutralizing background, Wigner posited that
when the electron density becomes sufficiently low, the system's
ground state should transform from a weakly interacting gas in which
the electrons are delocalized to a crystal in which the electrons are
strongly localized \cite{wigner34}. More precisely, Wigner conjectured
that at densities for which the average inter-electron distance much
exceeds the Bohr radius the ground state energy must be dominated by
the potential energy of the Coulombic repulsion of the electrons,
favoring a densely packed crystalline lattice. While Wigner
hypothesized that such a ground state would be given by a
body-centered cubic (BCC) lattice of electrons, it was Fuchs who
demonstrated numerically that another candidate in the form of a
close-packed lattice -- the face-centered cubic (FCC) lattice -- has a
slightly higher energy, confirming Wigner's claim among BCC, FCC and
simple cubic lattices \cite{fuchs35}.

In practice, the assumption of a uniform neutralizing background
(jellium) is quite unrealistic for typical solids. In fact, in his
calculation of the electrostatic energy of a lattice Fuchs considered
the energy of positive {\em ions} in the negative uniform background
of electrons \cite{fuchs35}. While nowadays such an assumption would
also be considered too crude for ordinary metals, it is actually quite
a reasonable one in an entirely different context, namely, that of
nuclear matter in the crust of neutron stars or other astrophysical
situations \cite{pethick95}. Perhaps the simplest model of such dense
nuclear matter would be a variant of Gamow's liquid drop model
specified in a macroscopically large volume with a fixed charge
density. Here the ultrarelativistic electrons provide the uniform
negative charge background, while nuclear matter arranges itself into
small proton-rich droplets corresponding to individual atomic nuclei.

An analysis of the above energy at low volume fraction of proton-rich
matter in three space dimensions reveals a limiting energy in which
the droplet shapes are determined by the competition of surface
tension with Coulombic self-energy of the droplets
\cite{kmn:cmp16,emmert20}, while the spatial arrangement of the
droplets minimizes the repulsive energy of point charges in a
neutralizing background \cite{frank26}. Recently, the problem
associated with the optimal droplet shape has been fully settled, with
the help of AI tools \cite{chodosh26}, yielding an implication that
all the droplets in the low density regime are asymptotically
identical balls that minimize the droplet energy per unit volume. In
that case the arrangement of the droplets reduces to that of the zero
temperature Coulomb gas in a uniform neutralizing background
\cite{serfaty}, which is mathematically equivalent to the Wigner
crystal.

Whether the above system forms a crystalline ground state, either as a
charged nuclear liquid or a zero temperature Coulomb gas is a very
difficult and fundamental unsolved problem in mathematical physics
\cite{alberti09,blanc15,serfaty,lewin22} (for a recent preprint by
OpenAI in two space dimensions, see
\cite{openai:abrikosov26}). Therefore, studies of this and other such
systems typically assume an a priori periodicity of the ground state
and then attempt to calculate the lowest energy state among a greatly
reduced configuration space of periodic lattices with a fixed
macroscopic particle density. For Coulombic systems, this amounts to
computing the Madelung constant of the charges
\cite{madelung18,borwein13,lewin19,lewin22}, a dimensionless quantity
characterizing the Coulombic interaction energy per unit volume in a
charge neutral periodic lattice. In the case of a single unit charge
per Wigner-Seitz cell $\Omega_L \subset \mathbb R^3$
\cite{kittel,lewin22} of a Bravais lattice
\begin{align}
  \label{eq:Lma}
  L = \{ m_1 \mathbf a_1 +m_2
  \mathbf a_2 + m_3 \mathbf a_3 \in \mathbb R^3 \ : \  (m_1, m_2, m_3)
  \in \mathbb Z^3 \},  
\end{align}
where $\mathbf a_1, \mathbf a_2, \mathbf a_3 \in \mathbb R^3$ are
three linearly independent lattice vectors, the Madelung constant is
most naturally defined as (using the normalization of
\cite{fuchs35,coldwell60})
\begin{align}
  \label{eq:Madelung}
  K_L := \frac{R_\mathrm{WS}}{2} \lim_{x \to 0} \left( v(x) - {1 \over
  |x|} \right), 
\end{align}
where $R_\mathrm{WS} := \sqrt[3]{3|\Omega_L| / (4 \pi)}$ is the
spherical Wigner-Seitz cell radius. Here the electrostatic potential
$v$ is the unique mean zero $L$-periodic solution of
\begin{align}
  \label{eq:vOmL}
  -\Delta v & = 4 \pi \Bigg( \sum_{p \in L} \delta_p -
              |\Omega_L|^{-1} \Bigg) \quad \text{in} 
              \quad \mathcal D'(\mathbb R^3),
\end{align}
where $\delta_p$ is the Dirac-delta centered at $p \in \mathbb R^3$
and $|\Omega_L|$ is the volume of the Wigner-Seitz cell, also referred
to as the covolume of the lattice $L$. The value of $K_L$ determines
the Coulombic energy $K_L e^2 / R_\mathrm{WS}$ of the Wigner crystal
electrons in Gaussian units per single charge $e$ and does not change
upon rescaling of $L$. Hence without loss of generality we may fix the
scale by choosing $|\Omega_L| = 1$, which we will do from now on. Note
that to within the normalization, $K_L$ coincides with the
renormalized energy of the Coulomb gas introduced in \cite{rougerie16}
(see also \cite{serfaty}). Wigner's conjecture would then state that
the BCC lattice minimizes the Madelung constant $K_L$ or,
equivalently, the renormalized Coulombic energy, among all Bravais
lattices $L$ of covolume one in three space dimensions.

To get a basic idea of the magnitude of the optimal Madelung constant,
one can make a simple approximation \cite{wigner38} that replaces the
Wigner-Seitz cell $\Omega_L$ of a tightly packed lattice with a ball
$B_{R_\mathrm{WS}}$ of the same volume, with Neumann boundary
condition on $\partial B_{R_\mathrm{WS}}$. A simple calculation then
yields the spherical approximation value to the Madelung constant:
\begin{align}
  \label{eq:KSW}
K_\mathrm{WS} = -0.9.  
\end{align}
This value turns out to be surprisingly close to the exact values of
the Madelung constants for the BCC and FCC lattices computed
numerically \cite{coldwell60,kozhberov21} (for certified numerical
values, see the following sections):
\begin{align}
  \label{eq:KBCCKFCC}
  K_{\mathrm{BCC}} \approx -0.895929255682,\qquad
  K_{\mathrm{FCC}}\approx -0.895873615195.
\end{align}
By inspection, $K_\mathrm{BCC}$ is the smallest among several lattice
types (see, e.g., the zero screening values in \cite{kozhberov18}),
making the conjecture among all Bravais lattices quite
plausible. Nevertheless, the chief difficulty in proving such a result
is a very small relative energy gap of $6.2 \times 10^{-5}$ between
the two lattices, which necessitates a very accurate evaluation of the
Coulombic energy.

It is well known that on a formal level the electrostatic potential
$v$ solving \eqref{eq:vOmL} may be thought of as the superposition of
the Coulombic potentials of point charges on the lattice minus that of
the neutralizing background. However, such sums are notoriously
ill-behaved \cite{borwein13}, so a suitable regularization is
necessary to make sense of such statements. Following the original
definition of Born \cite{born21}, for a covolume-one lattice $L$
define its Coulombic energy as
\begin{align}
  \label{eq:EL}
  E(L) := \lim_{\kappa \to 0^+} \left( \sum_{p \in L \backslash \{ 0
  \}} {e^{-\kappa |p|} \over |p|} - {4 \pi \over \kappa^2}
  \right) \sim \sum_{p \in L \backslash \{ 0
  \}} {1 \over |p|} , 
\end{align}
where the sum under the limit is an absolutely convergent series for
every $\kappa > 0$, representing the sum of the {\em screened} Coulomb
potentials of the charges on the lattice outside the origin that
formally corresponds to the (suitably renormalized) sum of the
electrostatic contributions of all the other charges. In fact, it is
easy to see that the limit in \eqref{eq:EL} is well defined, and by
Lemma \ref{lem:screened} we
have
\begin{align}
  \label{eq:KLEL}
  K_L = {R_\mathrm{WS} \over 2} E(L).
\end{align}
Alternatively, as is commonly done in the literature
\cite{borwein89,borwein13,lewin22}, the function $E(L)$ may be
obtained via an analytic continuation of the Epstein zeta function
\cite{epstein03, epstein06} (written in the convention of
\cite{luo26}):
\begin{align}
  \label{eq:Es}
  E(L, s) := \sum_{p \in L \backslash \{ 0
  \}} {1 \over |p|^{2s}}, \qquad \mathrm{Re} \, s > \frac32,
\end{align}
to $0 < s < \frac32$, which is known to be a meromorphic function of
$s$ with only a simple pole at $s = \frac32$ \cite{epstein03,
  epstein06,sarnak06}, by noting that \cite{borwein89,lewin19}
\begin{align}
  \label{eq:ELEsL}
  E(L) = E(L, \tfrac12).
\end{align}

Define the covolume-one FCC and BCC lattices \cite{luo26}:
\begin{equation}\label{eq:FB}
  \LF :=2^{-1/3}\bigl\{m\in\mathbb Z^3:\textstyle\sum_im_i\equiv0\ (2)\bigr\},
  \qquad
  \LB :=2^{-2/3}\bigl\{m\in\mathbb Z^3:m_1\equiv m_2\equiv m_3\
  (2)\bigr\}. 
\end{equation}
Our main result of this paper is the following theorem.

\begin{theorem}\label{thm:main}
  For every Bravais lattice $L\subset\mathbb R^3$ of covolume one we
  have
  \[
    E(L)\ \ge\ E(\LB),
  \]
  with equality if and only if $L$ is equal to $\LB$, up to an
  orthogonal transformation.
\end{theorem}

\noindent As we already noted, with $E(\LB) \approx -2.888461503$ the
margin to be established is
\begin{equation}\label{eq:gap}
  E(\LF)-E(\LB) \approx 1.7938403\times10^{-4},
\end{equation}
which makes the problem rather delicate.

Below we provide the description of our computer-assisted proof of
Theorem \ref{thm:main}, with the code that can be found in the
depositary \cite{zenodo26}. The code has been built and executed,
using AI (Claude Opus 5), and was then cross-validated by AI
(ChatGPT-6 Astra). The authors also manually verified all the
arguments and codes, and redid the verifying runs.

Our proof essentially follows that of Sarnak and Str\"ombergsson in
their proof that the FCC lattice uniquely minimizes the height
function of flat 3-tori, equivalently, minimizing the regular part
of the Epstein zeta function $E(L, s)$ at its pole $s = \frac32$ in
three space dimensions \cite{sarnak06}. We also use several analytic
tools from the work of Luo and Wei, who extended the conclusion of
Sarnak and Str\"ombergsson to the three-dimensional Epstein zeta
function $E(L, s)$ for all $s > \frac32$ in a recent preprint
\cite{luo26}. More precisely, as in \cite{sarnak06} we set up the
minimization problem by using the Iwasawa coordinates to represent all
the Bravais lattices in $\mathbb R^3$ as a 5-parameter family and then
minimize their Coulombic energy. We represent the latter, using Ewald
summation \cite{ewald21}, which gives the energy, up to an additive
constant, in the form of two rapidly convergent series of positive
terms:
\begin{align}
  \label{eq:ELewald}
  E(L)=-3 + \sum_{p\in L\setminus\{0\}}\frac{\erfc(\sqrt{\pi} |p|)}{|p|}
  +\frac1\pi\sum_{q\in L^{*}\setminus\{0\}}\frac{e^{-\pi |q|^2}}{|q|^2},
\end{align}
for any covolume-one lattice $L$, whose dual is $L^*$ (see the next
section for details).  Note that the minimization problem can be
equivalently posed in terms of the dual lattice $L^*$ instead, which
is what we actually do. Using just two terms in the Ewald sum, we
produce a short lattice vector cutoff that restricts the minimization
problem to a compact rectangular domain in $\mathbb R^5$. The proof
then obtains quantitative Hessian estimates in the neighborhood of the
FCC dual lattice, where we quantify the local minimality result of
Ennola for the FCC lattice with respect to $E(L, 1)$
\cite{ennola64}. We then evaluate the energy on a finite partition of
the remaining parameter space, using interval arithmetic and rigorous
bounds on the summands. Some care is required to ensure a sufficiently
high certified accuracy of these calculations that are carried out by
a computer.

Our paper is organized as follows. In Sec. \ref{sec:setup}, we present
the setup of the optimization problem by defining the energy with the
help of Iwasawa coordinates on the modified Grenier domain. In
Sec. \ref{sec:reduct-comp-rect}, we calculate the certified reference
value of the energy and reduce the minimization to a compact reduced
domain. In Sec. \ref{sec:local}, we carry out a quantitative
minimality study in a sufficiently small neighborhood of the reference
lattice. In Sec. \ref{sec:exterior}, we reduce the problem of
minimization outside the neighborhood of the reference lattice
characterized in Sec. \ref{sec:local} to estimating the energy on a
union of boxes with almost disjoint interior. Finally, in
Sec. \ref{sec:proof-theor} we construct a finite cover of the
minimization domain that allows us to certify the reference lattice as
the unique global energy minimizer, concluding the proof of Theorem
\ref{thm:main}. We also briefly discuss several possible directions of
related future research in Sec. \ref{sec:future}.

\section{Setup}
\label{sec:setup}

Throughout the paper, all vectors $p \in \mathbb R^3$ are assumed to
be column vectors, $|p|$ denotes their Euclidean norm and $|p|_\infty$
their infinity norm. For $g\in GL(3,\mathbb R)$ with $\det g=1$ put
$Y=g^{\mathsf T}g$ to be the Gram matrix of the columns of $g$, and
for $m \in \mathbb Z^3$ define $Y[m]=m^{\mathsf T}Ym = |gm|^2$. Then
$Y$ belongs to
\begin{align}
  \label{eq:P31}
  \PP=\{Y \in \mathbb R^{3 \times 3} \ : \ Y=Y^{\mathsf T}>0, \ \det
  Y=1\}, 
\end{align}
while the vectors $g m$ run over the covolume-one lattice $L$ formed
by the columns of $g$ as $m$ runs over $\mathbb Z^3$, see
\eqref{eq:Lma}. Moreover, every $Y \in \PP$ arises this way, $L$ is
uniquely determined by $Y$ up to rotations, and $Y$ is determined by
$L$ up to arithmetic congruence $Y \sim \gamma^{\mathsf T} Y \gamma$
for $\gamma \in GL(3, \mathbb Z)$. The dual lattice has Gram matrix
$Y^{-1}$. We write $\YF,\YB$ for Gram matrices of \eqref{eq:FB}.

We next define the representation of the energy in terms of Ewald
sums. As these formulas are not specific to the Coulomb kernel, we
introduce a general expression for the Epstein zeta function of a
lattice defined by its Gram matrix $Y \in \PP$ via the function (see
\cite[Sec. 4]{sarnak06}):
\begin{equation}\label{eq:H}
  \HH_s(Y) :=\sum_{m\in\mathbb Z^3\setminus\{0\}}
  \Bigl[G\bigl(s,\pi Y[m]\bigr)+G\bigl(\tfrac32-s,\pi
  Y^{-1}[m]\bigr)\Bigr], 
\end{equation}
where for every $a\in\mathbb R$ and $x>0$ the positive kernel $G(a,
x)$ is defined as
\begin{equation}\label{eq:G}
  G(a,x) :=\int_1^\infty e^{-xt}t^{a-1}\,dt.
\end{equation}
Then the following identity holds \cite[Eq. (23)]{sarnak06}: 
\begin{align}
  \label{eq:EsHs}
  \pi^{-s}\Gamma(s)E(Y,s)=\dfrac{1}{s-3/2}-\dfrac1s+\HH_s(Y),
\end{align}
where $\Gamma(s)$ is the gamma function, which, in particular, yields
\begin{align}
  \label{eq:ELHH12}
  E(L) = \HH_{1/2}(Y_L) - 3,
\end{align}
where $Y_L$ is the Gram matrix of the covolume-one lattice $L$. Thus,
minimizing $E(L)$ among all such $L$ is equivalent to minimizing
$\HH_{1/2}(Y)$ among all $Y \in \PP$. Also, from the duality relation
\cite[Eq. (2)]{sarnak06}
\begin{align}
  \label{eq:Hsdual}
  \HH_s(Y) = \HH_{3/2-s}(Y^{-1})
\end{align}
for every $Y \in \PP$ follows that minimizing $E(L)$ in $L$ is
equivalent to minimizing $\HH_1(Y)$ among all $Y \in \PP$, and the
statement of Theorem \ref{thm:main} that $E(L)$ attains its minimum
exactly at $L = \LB$ is equivalent to
\begin{equation}\label{eq:target}
  \HH_1(Y)\ \ge\ \HH_1(\YF)\quad\text{for all} \  Y \in \PP, \
  \text{with equality only if } Y \text{ is congruent to} \ \YF. 
\end{equation}
From now, we refer to $\HH_1(Y)$ as the energy to be minimized among
all $Y \in \PP$. Notice that 
\begin{align}
  \label{eq:G112}
  G\left( \tfrac12, x \right) = {\sqrt{\pi} \erfc \left( \sqrt{x}
  \hspace{0.5mm} \right) \over \sqrt{x}}, \qquad G(1,
  x) = {e^{-x} \over x},
\end{align}
so together with \eqref{eq:H} and \eqref{eq:ELHH12} this gives
\eqref{eq:ELewald}. 

We now introduce the parametrization of the matrices in $\PP$ by the
Iwasawa coordinates $\xi=(y_1,y_2,a,b,c)$, which form a
five-dimensional parameter space\cite[Sec. 4]{sarnak06}:
\begin{equation}\label{eq:iwasawa}
  Y=N^{\mathsf T}DN,\qquad
  N=\begin{pmatrix}1&a&b\\0&1&c\\0&0&1\end{pmatrix},\qquad
  D=\operatorname{diag}(d_1,d_2,d_3),
\end{equation}
\begin{align}
  \label{eq:d123}
  d_1=(y_1^2y_2)^{1/3}, \qquad d_2=(y_2/y_1)^{1/3},  \qquad d_3= (y_1 y_2^2)^{-1/3},   
\end{align}
so that $\det Y=1$ identically. This
defines a diffeomorphism from
$\mathbb R^+ \times \mathbb R^+ \times \mathbb R \times \mathbb R
\times \mathbb R$ onto $\PP$. We then use the modified Grenier
fundamental domain \cite{grenier88} that is specified by
$0\le a\le\frac12$, $0\le b\le1$, $0\le c\le\frac12$ together with
\begin{equation}\label{eq:red}
\begin{aligned}
  1&\le(1+a-b)^2+y_1^{-1}\bigl((1-c)^2+y_2^{-1}\bigr), &\quad
  1&\le(a-b)^2+y_1^{-1}\bigl((1-c)^2+y_2^{-1}\bigr),\\
  1&\le a^2+y_1^{-1}, &\quad
  1&\le b^2+y_1^{-1}(c^2+y_2^{-1}),\\
  1&\le(b-1)^2+y_1^{-1}(c^2+y_2^{-1}), &\quad
  1&\le c^2+y_2^{-1}.
\end{aligned}
\end{equation}
We call $\xi$, or the corresponding $Y_\xi$, reduced if it satisfies
the coordinate constraints on $a, b, c$ together with
\eqref{eq:red}. By \cite[Sec.~4]{sarnak06} every $Y\in\PP$ is
$GL(3,\mathbb Z)$-congruent to at least one reduced $Y_\xi$, and
$\HH_1$ is invariant under such congruences, so it suffices to
minimize over reduced representatives.  Note that
\begin{align}
  \label{eq:Y100}
  Y[(1, 0, 0)] = d_1.
\end{align}
Also note that in this domain the FCC lattice has a unique
representative \cite{sarnak06}
\begin{align}
  \label{eq:xiF}
  \xi_\mathrm{F} = \left( \frac43, \frac98, \frac12, \frac12, \frac13 \right),
\end{align}
for which
\begin{align}
  \label{eq:YF}
  \YF = 2^{-2/3} \begin{pmatrix}2&1&1\\1&2&1\\1&1&2\end{pmatrix}. 
\end{align}

Lastly, for $m=(i,j,k) \in \mathbb Z^3$ the Iwasawa form
\eqref{eq:iwasawa} and direct inversion give
\begin{align}
  Y[m]
  & =d_1(i+aj+bk)^2+d_2(j+ck)^2+d_3k^2, \label{eq:Ym} \\
  Y^{-1}[m]
  &
    =\frac{i^2}{d_1}+\frac{(j-ai)^2}{d_2}+\frac{(k-cj+(ac-b)i)^2}{d_3},
    \label{eq:YYm} 
\end{align}
both weighted sums of squares and, hence, enclosable tightly in the
interval arithmetic to obtain outward-rounded intervals of \ieee{}
binary64 numbers. The computation of $Y[m]$ and $Y^{-1}[m]$ involves
extraction of cube roots in \eqref{eq:d123}, which is done by first
using library seeds and then proving them in interval arithmetics by
cubing the result with directed rounding. The same seed-and-prove
procedure is used for the square roots appearing in the reduced-domain
enclosure and in the Frobenius norms of the Hessian tail estimates. In
computing the Hessian, derivatives are obtained throughout by
second-order forward automatic differentiation: the value, gradient
and Hessian are propagated together through each arithmetic operation,
with the derivatives of $x^{-1}$, $x^{1/3}$ and $G(a,\cdot)$ supplied
in closed form, so that no gradient or Hessian of \eqref{eq:Ym} or
\eqref{eq:YYm} is ever transcribed.

\section{Reduction to a compact rectangle}
\label{sec:reduct-comp-rect}

We now proceed to the proof of \eqref{eq:target}. We start by
observing that the modified Grenier domain defined above is not
compact, since one can choose, for example, $a = b = c = 0$ and
$y_1 = y_2 = d \to 0^+$, with the limit not in the domain. This,
however, produces $d_1 = d \to 0$, and \eqref{eq:Y100} then implies
that the corresponding lattice degenerates, causing the energy to
explode. Such a behavior, however, may be excluded by an energy
comparison.

First, we need a tight bound on the energy of $\YF$.

\begin{lemma}
  \label{lem:H1YF}
  We have
  \begin{align}
    \label{eq:H1YF}
    \HH_1(\YF) \in [0.1115384969460437,\,0.11153849694607266]. 
  \end{align}
\end{lemma}

\begin{proof}
  To obtain a precise bracketing of the value of $\HH_1(\YF)$, we use
  the fact that the Ewald sum \eqref{eq:H} is a sum of positive terms
  and, therefore, a lower bound follows by truncating the sum to a
  finite number of shells around the origin. An upper bound is then
  obtained by estimating the rapidly converging tail of the series in
  \eqref{eq:H} outside those shells. To achieve the desired accuracy,
  it is sufficient to consider the shells satisfying
  $|m|_\infty \leq 8$ and combine every pair $\pm m$ into a single
  summand. The formula is built from the exponential function in the
  definition of $G(1, \pi Y[m])$, which is evaluated by repeated
  halving of the argument (which carries no error in binary
  arithmetic) to reduce it to the interval $[0, \frac12]$ with a
  geometric tail bound and subsequent repeated squaring. On the other
  hand, as in \cite{luo26} the function $G(\frac12, \pi Y^{-1}[m])$
  can be bracketed \cite[Ch. III]{stieltjes94} between the two
  consecutive convergents of the Stieltjes continued fraction
  representation \cite[Eq.~(12.6.15b)]{cuyt} of
  \begin{equation}\label{eq:Ghalfcf}
    G\!\left(\tfrac12,x\right)
    \;=\;\cfrac{e^{-x}}{x+\cfrac{1/2}{1+\cfrac{1}{x+\cfrac{3/2}{1
            +\cfrac{2}{x+\cfrac{5/2}{1+\ddots}}}}}}, 
    \qquad x>0 , 
  \end{equation}
  at orders 39 and 40. Numerically at order 39/40 the bracket is
  already tight: for $Y = \YF$ we have $x = \pi Y^{-1}[m] > 3.7$,
  where the bracket is below $10^{-14}$. All the algebra is done in
  outward-rounding interval arithmetic.

  Finally, to estimate the tail we note that shell
  $|m|_\infty = n \in \mathbb N$ has $24n^2+2 \le 26n^2$ vectors and
  $Y[m] \ge \mu n^2$, where $\mu := 1/\operatorname{tr}(Y^{-1})$ is a
  lower bound for the smallest eigenvalue of $Y$, while
  $Y^{-1}[m] \ge \mu^* n^2$, where $\mu^* := 1/\operatorname{tr}(Y)$
  is a lower bound for the smallest eigenvalue of $Y^{-1}$,
  respectively. Since $G(a, x)$ is increasing in $a$ and decreasing in
  $x$ for all $a \in \mathbb R$ and $x > 0$, we can estimate the
  contribution $S_n$ of that shell by
  \begin{multline}
    \label{eq:Gtail}
    S_n = \sum_{|m|_\infty = n} \left[ G(1, \pi Y[m]) + G\left(
        \tfrac12, \pi Y^{-1}[m] \right) \right] \leq 26 n^2 \left[
      G(1, \pi \mu n^2) + G(1, \pi \mu^* n^2) \right]
    \\
    \leq {26 r_N^{n-N} e^{-\pi \mu N^2} \over \pi \mu} + {26 r_N^{n-N}
      e^{-\pi \mu^* N^2} \over \pi \mu^*}, \qquad r_N = e^{-\pi (2 N +
      1) \min (\mu, \mu^*)},
  \end{multline}
  for all $n \geq N \in \mathbb N$. Therefore, the contribution of all
  the shells for $|m|_\infty > 8$ is bounded above by
  \begin{align}
    \label{eq:Gtail2}
    \sum_{n=N}^\infty S_n \leq  {26 e^{-\pi \mu N^2} \over \pi (1 -
    r_N) \mu} + {26  e^{-\pi \mu^* N^2} \over \pi (1 - r_N) \mu^*},
    \qquad N = 9.
  \end{align}
  For $Y = \YF$ we have, computing the traces of $\YF$ and $\YF^{-1}$
  from \eqref{eq:Ym} and \eqref{eq:YYm} in interval arithmetic,
  $\mu \ge 0.27998245553219342$ and $\mu^* \ge 0.26456684199469938$,
  yielding an $1.85\times10^{-28}$ upper bound on the tail. The final
  interval is the result of combining this estimate with the
  evaluation of the finite sum in interval arithmetic by the code.
\end{proof}

Note that the same procedure for the BCC lattice, whose unique
representative in the reduced domain is
$\xi_{\mathrm B} = (\tfrac98,\tfrac43,\tfrac13,\tfrac23,\tfrac12)$,
gives
$\HH_1(Y_{\mathrm B}) \in [0.11171788097966455,\
0.11171788097968673]$, which is disjoint and above the interval in
Lemma \ref{lem:H1YF}.

The next lemma provides a lower bound for $d_1$ for all $Y \in \PP$
whose energy does not exceed that of $\YF$. Note that by
\eqref{eq:Y100} it rules out the presence of a short lattice vector.

\begin{lemma}
  \label{lem:cusp}
  If $Y \in \PP$ and $d_1 < \frac23$, where $d_1$ is as in Section
  \ref{sec:setup}, then $\HH_1(Y)>\HH_1(\YF)$, with the margin of at
  least $6.056\times10^{-3}$.
\end{lemma}

\begin{proof}
  By positivity of the summands in \eqref{eq:H} and with the help of
  \eqref{eq:G112} and \eqref{eq:Y100}, we have
  $\HH_1(Y)\ge 2G(1,\pi Y[(1, 0, 0)] )=2e^{-\pi d_1}/(\pi d_1)$, which
  is strictly decreasing in $d_1$ and is hence bounded below by its
  value at $d_1 = 2/3$ for all $d_1 \le 2/3$.  By direct computation
  of the exponential function as in Lemma \ref{lem:H1YF} we obtain
  \begin{align}
    \label{eq:Gd1}
    2G(1,2\pi/3)\ge 0.1175945368946040. 
  \end{align}
  Comparing this with the bound in Lemma~\ref{lem:H1YF} yields the
  claim, with a certified separation of at least
  $6.0560399\times10^{-3}$.
\end{proof}

Lemma~\ref{lem:cusp} restricts attention in the minimization of
$\HH_1$ to $d_1\ge2/3$. The third and sixth inequalities of
\eqref{eq:red}, together with the upper bounds on $a$ and $c$ give
$y_1,y_2\le4/3$. Combining this with
$d_1^3=y_1^2y_2 \geq {8 \over 27}$ then yields
$y_1\ge {\sqrt2 \over 3}$ and $y_2\ge\frac16$. Hence every remaining
reduced representative of $Y \in \PP$ whose energy does not exceed
that of $\YF$ lies in the compact rectangle
\begin{equation}\label{eq:rect}
  \mathcal
  K=\Bigl[\tfrac{\sqrt2}{3},\tfrac43\Bigr]\times\Bigl[\tfrac16,\tfrac43\Bigr]  
  \times\Bigl[0,\tfrac12\Bigr]\times[0,1]\times\Bigl[0,\tfrac12\Bigr].
\end{equation}
Since $\mathcal K$ is compact and $\HH_1$ is continuous in the Iwasawa
coordinates, the global minimum is attained; it remains to locate it.

On $\mathcal K$ one has $\frac13 \le d_1 \le \frac43$, $\frac12 \le
d_2\le \sqrt{2}$
and $\frac34 \le d_3\le 3 \sqrt{2}$. Consequently, for every nonzero
$m\in\mathbb Z^3$,
\begin{equation}\label{eq:qbound}
  Y[m]\ \ge\ \frac13,\qquad Y^{-1}[m]\ \ge\ {\sqrt{2} \over 6}
  \qquad(Y\in\mathcal K). 
\end{equation}
These bounds guarantee that no kernel is ever evaluated near its
singularity.

\section{Local analysis at the reference lattice}\label{sec:local}

We now discuss the behavior of the energy around the reference
FCC lattice. As was shown by Ennola \cite{ennola64}, FCC lattice is a
strict local minimizer of $\HH_1$ (for criticality, see also
\cite{betermin19}). However, Ennola's result does not quantify the
region of strict local minimality around FCC, which is what we need
to exclude a finite neighborhood of $\YF$ in our exhaustive search of
the minimizer of $\HH_1(Y)$. The latter is the contents of the
following proposition.

\begin{proposition}
  \label{prop:ball}
  Let $\delta_0 = 2 \times 10^{-4}$ and let
  $R_0 := \prod_{j=1}^5 [\xi_{\mathrm{F},j}-\delta_0,
  \xi_{\mathrm{F},j}+\delta_0] \subset \mathbb R^5$, where
  $\xi_{\mathrm F}$ is defined in \eqref{eq:xiF}, be a rectangle of
  Iwasawa coordinates. Then the Hessian of $\HH_1$ is positive
  definite for every $Y_\xi$ such that $\xi \in R_0$ and, as a
  consequence, $\HH_1(Y_\xi)\ge\HH_1(\YF)$ for all $\xi\in R_0$, with
  equality only at $\xi_{\mathrm F}$.
\end{proposition}

\begin{proof}
  Write $\HH_1(Y_\xi)=P(\xi)+T(\xi)$, where
\begin{align}\label{eq:ball-split}
  P(\xi) &:=\sum_{0<|m|_\infty\le2}
           \Bigl[G(1,\pi Y_\xi[m])+G(\tfrac12,\pi Y_\xi^{-1}[m])\Bigr],
  & T(\xi) &:=\HH_1(Y_\xi)-P(\xi) .
\end{align}
By the local uniform convergence of \eqref{eq:H} and of its
derivatives of any fixed order on compact subsets of $\PP$, both $P$
and $T$ are $C^2$ on a neighborhood of $R_0$ and may be
differentiated termwise.

\emph{Step 0: recurrence and an elementary kernel bound.} Integration
by parts in \eqref{eq:G} gives, for every $a\in\mathbb R$ and $x>0$,
\begin{align}\label{eq:Grec}
  G(a+1,x) &= \frac{e^{-x}+a\,G(a,x)}{x},
\end{align}
so the orders $a+1$ and $a+2$ required in \eqref{eq:ball-d2} below are
obtained from $G(1,x)={e^{-x} \over x}$ and $G(\tfrac12,x)$ by up to
two applications of \eqref{eq:Grec}. In particular, we have
$G(3,x) = \frac{e^{-x}}{x}\Bigl(1+\frac2x+\frac2{x^2}\Bigr)$. Since
$G(a,x)$ is increasing in $a$, for every $a\le3$ and $x \ge 1$ there
holds
\begin{align}\label{eq:ball-Gall}
  G(a,x) &\le \frac{5e^{-x}}{x}.
\end{align}

\emph{Step 1: the omitted shells.} Set
\begin{align}\label{eq:ball-tau}
  \tau &:=\max_{1\le j,k\le5}\ \sup_{\xi\in R_0}\
    \sum_{|m|_\infty>2}
    \Bigl|\partial^2_{jk}\bigl[G(1,\pi Y_\xi[m])
      +G(\tfrac12,\pi Y_\xi^{-1}[m])\bigr]\Bigr| .
\end{align}
Differentiating twice and using $\partial_xG(a,x)=-G(a+1,x)$, we get
the terms of the type
\begin{align}\label{eq:ball-d2}
  \partial^2_{jk}G(a,\pi Q)
  &=\pi^2G(a+2,\pi Q)\,\partial_jQ\,\partial_kQ-\pi G(a+1,\pi
    Q)\,\partial^2_{jk}Q  
\end{align}
in \eqref{eq:ball-tau}, for $Q=Y_\xi[m]$ when $a=1$ and for
$Q=Y_\xi^{-1}[m]$ when $a=\tfrac12$. Note that only the orders $a\le3$
of $G(a, \pi Q)$ occur. The shells $3\le|m|_\infty\le8$ are bounded by
evaluating the same second-order jets over $R_0$ and accumulating the
absolute values entrywise in interval arithmetic. For $|m|_\infty=n>8$
put $\mu:=1/\operatorname{tr}(Y_\xi^{-1})$ and
$\mu^*:=1/\operatorname{tr}(Y_\xi)$, lower bounds for the smallest
eigenvalues of $Y_\xi$ and $Y_\xi^{-1}$, respectively. An evaluation
over $R_0$ yields $\mu \in [0.279890149,\ 0.280074769]$. Hence
$Y_\xi[m] \ge\mu n^2 \geq 1$ on $R_0$, and with $K_1 \leq 2.360$ and
$K_2 \leq 2.521$ denoting Frobenius bounds for $\partial_jY_\xi$ and
$\partial^2_{jk}Y_\xi$ on $R_0$ obtained by automatic differentiation
of \eqref{eq:Ym} and evaluating the bounds over $R_0$ in interval
arithmetic, we have
\begin{align}\label{eq:ball-far}
  |\partial_jQ| &\le K_1\|m\|_2^2\le 3K_1n^2, &
  |\partial^2_{jk}Q| &\le 3K_2n^2 ,
\end{align}
since $\|m\|_2^2\le3n^2$ on the shell. Combining \eqref{eq:ball-Gall},
\eqref{eq:ball-d2} and \eqref{eq:ball-far} with
$\#\{|m|_\infty=n\} \le 26n^2$, the shell contributes at most
\begin{align}\label{eq:ball-shell}
  S_n \leq \frac{26\,e^{-\pi\mu n^2}}{\mu}\Bigl(45\pi K_1^2n^4+15K_2n^2\Bigr).
\end{align}
Analogously, we compute $\mu^* \in [0.264483750,\ 0.264649949]$, and
with $K_1^* \leq 2.102$ and $K_2^* \leq 2.978$ the respective
Frobenius bounds for $\partial_jY_\xi^{-1}$ and
$\partial^2_{jk}Y_\xi^{-1}$, for $Q = Y^{-1}[m] \geq \mu^* n^2 \geq 1$
on $R_0$ each summand has the form $Cn^pe^{-cn^2}$ with $p\le4$, so
consecutive ratios of those terms are
$\left( 1+n^{-1} \right)^pe^{-c(2n+1)}$. These ratios are decreasing
in $n$, hence their value at $n=9$ and $p = 4$ majorizes all later
ones and the series is summed by comparison with a geometric one as in
the proof of Lemma \ref{lem:H1YF}. The resulting far-field
contribution is below $10^{-20}$, whence from the dominant
contribution of the $n = 3$ shell we get
\begin{align}\label{eq:ball-tauval}
  \tau &\le 7.977613\times10^{-9},
\end{align}
and $\bigl|\partial^2_{jk}T(\xi)\bigr|\le\tau$ for all $\xi\in R_0$
and all $j,k$.

\emph{Step 2: an enclosure of the Hessian.} Evaluating the
second-order jets of $P$ over $R_0$ in interval arithmetic --- that
is, with each Iwasawa coordinate replaced by the interval
$[\xi_{\mathrm F,j}-\delta_0,\ \xi_{\mathrm F,j}+\delta_0]$ --- yields
intervals $\bigl[\underline p_{jk},\overline p_{jk}\bigr]$ containing
$\partial^2_{jk}P(\xi)$ simultaneously for every $\xi\in
R_0$. Widening each entry by the bound \eqref{eq:ball-tauval} of
Step~1,
\begin{align}\label{eq:ball-A}
  A_{jk} &:=\bigl[\underline p_{jk}-\tau,\ \overline p_{jk}+\tau\bigr],
  \qquad 1\le j,k\le 5 ,
\end{align}
and recalling $\HH_1(Y_\xi)=P(\xi)+T(\xi)$, we obtain
$\partial^2_{jk}\HH_1(\xi)\in A_{jk}$ for every $\xi\in R_0$. Thus the
symmetric matrix $\nabla^2\HH_1(\xi)$ belongs to the interval matrix
$A$ at each point of $R_0$.

\emph{Step 3: positive definiteness.} If $A$ were a fixed symmetric
positive definite matrix, we could determine its positivity by
carrying out the square-root-free Cholesky factorization
$A = LDL^{\mathsf T}$, where $L$ is the unit lower triangular matrix
and $D$ is a diagonal matrix \cite[Sec. 4.1.2]{golub}. However, our
$A$ is an interval matrix, hence to determine the bounds on the
respective pivots we apply the $LDL^{\mathsf T}$ recursion:
\begin{align}\label{eq:ball-ldl}
  d_j &=A_{jj}-\sum_{k<j}L_{jk}^2\,d_k, &
  L_{ij} &={1 \over d_j} \Bigl(A_{ij}-\sum_{k<j}L_{ik}L_{jk}\,d_k\Bigr)
  \quad (i>j),
\end{align}
in interval arithmetic. The computation returns
\begin{align}\label{eq:ball-pivots}
  d_1&\in[0.084881,\,0.088247], & d_2&\in[0.095808,\,0.114909],\nonumber\\
  d_3&\in[0.367913,\,0.379133], & d_4&\in[0.232841,\,0.264387],\nonumber\\
  d_5&\in[0.076444,\,0.157416], &&
\end{align}
so every pivot is bounded away from zero, with $\tau$ smaller by a
factor of $10^7$. In particular, no division in \eqref{eq:ball-ldl} is
by an interval containing $0$, and inclusion monotonicity of the
interval operations gives, by induction on $j$, that for \emph{every}
symmetric $H\in A$ the exact recursion is well defined and its
quantities lie in the computed intervals. Each exact pivot therefore
satisfies $d_j\ge \min_j \underline{d}_j \geq 0.0764$, so
$H=LDL^{\mathsf T}$ with $L$ unit lower triangular and $D$ diagonal
with positive entries, whence for $x\ne0$
\begin{align}\label{eq:ball-quadform}
  x^{\mathsf T}Hx &=y^{\mathsf T}Dy=\sum_{j=1}^5 d_j\,y_j^2>0,
  \qquad y:=L^{\mathsf T}x\ne0 ,
\end{align}
since $L^{\mathsf T}$ is invertible. By Step~2 this applies to
$H=\nabla^2\HH_1(\xi)$ for every $\xi\in R_0$, which proves the first
assertion.

\emph{Step 4: the comparison.} The rectangle $R_0$ is convex and
contains $\xi_{\mathrm F}$.  Fix $\xi\in R_0$ and put
$v:=\xi-\xi_{\mathrm F}$, so that $\xi_{\mathrm F}+tv\in R_0$ for all
$t\in[0,1]$. Taylor's formula with integral remainder, together with
the stationarity $\nabla\HH_1(\YF)=0$, gives
\begin{align}\label{eq:ball-taylor}
  \HH_1(Y_\xi)-\HH_1(\YF)
  &=\int_0^1(1-t)\,v^{\mathsf T}\nabla^2\HH_1\bigl(\xi_{\mathrm
    F}+tv\bigr)\,v\,dt . 
\end{align}
By Step~3 the integrand is nonnegative on $[0,1]$, and strictly
positive for $v\ne0$ and all $t \in (0,1)$. Hence
$\HH_1(Y_\xi)\ge\HH_1(\YF)$ throughout $R_0$, with equality only at
$\xi=\xi_{\mathrm F}$.
\end{proof}

\section{The exterior comparison}\label{sec:exterior}

Let $\mathcal V$ contain one representative of each pair $\{m,-m\}$ in
$([-2,2]^3\cap\mathbb Z^3)\setminus\{0\}$, so $|\mathcal V|=62$, and
\begin{align}
  \label{eq:PYxi}
  P(\xi)=2\sum_{m\in\mathcal V}\bigl[G(1,\pi Y_\xi[m])+G(\tfrac12,\pi
  Y_\xi^{-1}[m])\bigr]
\end{align}
be as in the proof of Proposition \ref{prop:ball}.  Positivity of the
summands gives $\HH_1(Y_\xi)\ge P(\xi)$ for every
$\xi \in \mathbb R_+^2 \times \mathbb R^3$, with no correction term:
the omitted part may simply be discarded. Being a finite sum of smooth
functions of $\xi$, $P$ is $C^2$ on $\mathbb R_+^2\times\mathbb R^3$.

\begin{proposition}\label{prop:cert}
  Let
  $B=\prod_{j=1}^5[c_j-r_j,\,c_j+r_j]\subset\mathbb R_+^2\times\mathbb
  R^3$ be a box of Iwasawa coordinates with center
  $c \in \mathbb R_+^2 \times \mathbb R^3$ and coordinate radii
  $r\in\mathbb R_+^5$, and let $H$ be an interval enclosure of the
  Hessian of $P$ over $B$, with $\underline{H_{ii}}$ and
  $\overline{|H_{ij}|}$ denoting the corresponding bounds. Put
  \begin{align}\label{eq:cert-kappa}
    \kappa_i &:=\underline{H_{ii}}-\sum_{j\ne i}\overline{|H_{ij}|}\,\frac{r_j}{r_i},
    \qquad i=1,\dots,5 .
  \end{align}
  Then
  \begin{align}\label{eq:cert}
    \inf_B P
    &\ \ge\ P(c)-\sum_{j=1}^5 r_j\bigl|\partial_jP(c)\bigr|
      +\tfrac12\sum_{i=1}^5\min(0,\kappa_i)\,r_i^2 .
  \end{align}
\end{proposition}

\begin{proof}
  Every point of
  $B$ is $c+v$ with $|v_j|\le r_j$. By Taylor's theorem with Lagrange
  remainder there is a $w$ on the segment from $c$ to $c+v$ with
  \begin{align}\label{eq:cert-taylor}
    P(c+v) &= P(c)+\nabla P(c)\cdot v+\tfrac12\,v^{\mathsf T}\nabla^2P(w)\,v .
  \end{align}
  The gradient is evaluated at the single point $c$, whereas $w$ is unknown and the
  Hessian must therefore be estimated through its enclosure $H$ over all of $B$. The linear
  term is bounded directly from $|v_j|\le r_j$:
  \begin{align}\label{eq:cert-lin}
    \nabla P(c)\cdot v &\ \ge\ -\sum_{j=1}^5 r_j\bigl|\partial_jP(c)\bigr| .
  \end{align}

  For the quadratic term, separating the diagonal from the off-diagonal entries and using
  $\nabla^2P(w)\in H$ gives
  \begin{align}\label{eq:cert-quad0}
    v^{\mathsf T}\nabla^2P(w)\,v
    &\ \ge\ \sum_{i=1}^5\underline{H_{ii}}\,v_i^2
      -\sum_{i\ne j}\overline{|H_{ij}|}\,|v_i|\,|v_j| .
  \end{align}
  We now apply Young's inequality with the weights $r_j/r_i$ supplied by the box itself,
  \begin{align}\label{eq:cert-young}
    |v_i|\,|v_j| &\ \le\ \tfrac12\Bigl(\tfrac{r_j}{r_i}v_i^2+\tfrac{r_i}{r_j}v_j^2\Bigr) ,
  \end{align}
  to each cross term; this is where $r_i>0$ is used. Summing over $i\ne j$ and using the
  symmetry $H_{ij}=H_{ji}$, the two halves of \eqref{eq:cert-young} coincide after
  relabelling $i\leftrightarrow j$, so the cross terms contribute
  $\sum_i v_i^2\sum_{j\ne i}\overline{|H_{ij}|}\,r_j/r_i$ and, by \eqref{eq:cert-kappa},
  \begin{align}\label{eq:cert-kappa2}
    v^{\mathsf T}\nabla^2P(w)\,v &\ \ge\ \sum_{i=1}^5\kappa_i\,v_i^2 .
  \end{align}
  A term with $\kappa_i\ge0$ is nonnegative and may be discarded,
  while a term with $\kappa_i<0$ satisfies
  $\kappa_i v_i^2\ge\kappa_i r_i^2$; in both cases
  $\kappa_i v_i^2\ge\min(0,\kappa_i)r_i^2$. Combining this with
  \eqref{eq:cert-taylor} and \eqref{eq:cert-lin}, the right-hand side
  of \eqref{eq:cert} bounds $P(c+v)$ from below for every admissible
  $v$, and taking the infimum over $v$ proves the claim.
\end{proof}

In the implementation, the center $c$ is a floating-point vector
chosen inside the original box $B$, and the radii are rounded upward
so that $|\xi_j-c_j| \leq r_j$ for every $\xi \in B$. The Hessian
enclosure is evaluated on $B$.  The preceding proof applies with these
possibly enlarged radii, since the segment joining $c$ to any $\xi$ in
$B$ remains in $B$; an exact midpoint and an equality
$B = \prod_{j=1}^5 [c_j-r_j,c_j+r_j]$ are not required.

\section{Proof of Theorem \ref{thm:main}}\label{sec:proof-theor}

We now describe the sequential subdivision of the compact set
$\mathcal K \subset \mathbb R_+^2 \times \mathbb R^3$ into a finite
set of rectangles by bisecting the parent at the midpoint of its
widest coordinate, with both closed box children retained, so their
union is the parent. The endpoints of \eqref{eq:rect} are rounded
outward to binary64 numbers; the resulting rectangle contains
\eqref{eq:rect}, so covering it suffices. A box $B\subseteq\mathcal K$
is terminal if one of the following is verified, in this order:
\begin{description}
\item[\textnormal{(R)}] one of the six inequalities \eqref{eq:red} has
  upper endpoint below $1$ throughout $B$, so $B$ contains no reduced
  representative;
\item[\textnormal{(C)}] $d_1<\frac23$ throughout $B$, so
  Lemma~\ref{lem:cusp} applies;
\item[\textnormal{(B)}] $B\subseteq R_0$ from Proposition
  \ref{prop:ball};
\item[\textnormal{(E)}] the right-hand side of \eqref{eq:cert} exceeds
  the upper endpoint of the certified enclosure of $\HH_1(\YF)$ in
  Lemma \ref{lem:H1YF}.
\end{description}

By induction the terminal boxes cover $\mathcal{K}$. Each of
\textnormal{(R)}, \textnormal{(C)}, \textnormal{(B)}, \textnormal{(E)}
is decidable in interval arithmetic, and the process is run until no
box remains unresolved. That it does so after finitely many
subdivisions, and with how many boxes, is the computational content of
the proof, as summarized in the following proposition.

\begin{proposition}\label{prop:cover}
  The bisection described above terminates. It produces a covering of
  the rectangle \eqref{eq:rect} by $46\, 414$ closed boxes, on each of
  which one of \textnormal{(R)}, \textnormal{(C)}, \textnormal{(B)},
  \textnormal{(E)} is verified, with no unresolved box:
  \[
  \begin{array}{lr}
    \toprule
    \text{terminal boxes} & 46\, 414\\
    \midrule
    \text{(R) reduction-empty} & 6\, 967\\
    \text{(C) short lattice vector} & 2\, 444\\
    \text{(B) small FCC neighborhood} & 45\\
    \text{(E) second-order comparison} & 36\, 958\\
    \text{unresolved} & 0\\
    \midrule
    \text{smallest margin} & 5.10236\times10^{-12} \\
    \bottomrule
  \end{array}
  \]
\end{proposition}

Building the cover takes under 5 minutes, and the certification run
takes about two minutes on one core of a current laptop computer.  The
subdivision is recorded as a sequence of split instructions, each
naming the coordinate that was bisected and the midpoint used. The
certifying program takes this record as untrusted input: it constructs
the root rectangle itself, and for every instruction checks that the
midpoint lies strictly inside the current box before forming both
children, so that the union of the boxes it builds is the root at
every stage. It then re-tests each terminal box against
\textnormal{(R)}, \textnormal{(C)}, \textnormal{(B)}, \textnormal{(E)}
in interval arithmetic, and verifies that the record is exhausted with
no box left open and that the number of terminal boxes exceeds the
number of splits by exactly one. A corrupted record is therefore
rejected rather than believed; in particular the covering of
Proposition~\ref{prop:cover} does not depend on the program that
produced it.

\begin{proof}[Proof of Theorem~\ref{thm:main}]
  Let $\xi\in\mathcal K$ be a reduced representative. By
  Proposition~\ref{prop:cover}, $\xi$ lies in some terminal box
  $B$. If $B$ is of type \textnormal{(R)} then $B$ contains no reduced
  representative, contradicting the fact that $\xi$ is reduced. If $B$
  is of type \textnormal{(C)} then Lemma~\ref{lem:cusp} gives
  $\HH_1(Y_\xi)>\HH_1(\YF)$. If $B$ is of type \textnormal{(B)} then
  $B\subseteq R_0$, and Proposition~\ref{prop:ball} gives
  $\HH_1(Y_\xi)\ge\HH_1(\YF)$, with equality only at
  $\xi=\xi_{\mathrm F}$. Finally, if $B$ is of type \textnormal{(E)}
  then, by positivity of the summands and Proposition~\ref{prop:cert},
\[
  \HH_1(Y_\xi)\ \ge\ P(\xi)\ \ge\ \inf_B P\ >\ \HH_1(\YF).
\]
Hence
$\HH_1(Y_\xi)\ge\HH_1(\YF)$ for every reduced $\xi\in\mathcal K$, with
equality only at $\xi_{\mathrm F}$; by the passage from the reduced
domain established in Lemma~\ref{lem:cusp} and \eqref{eq:rect} this is
a global minimum over $\PP$. The theorem now follows by the passage to
the dual lattice.
\end{proof}

\section{Further directions}\label{sec:future}

We finish by discussing several possible directions to extend our
work.

\paragraph{Other Riesz exponents.}
An immediate extension is to minimize the analytically continued
Epstein zeta function $E(L,s)$ for exponents other than the Coulomb
value $s=1/2$ in three space dimensions. In the present convention,
where the interaction is $|x|^{-2s}$, the expected phase diagram among
covolume-one Bravais lattices is\cite[Eq. (44)]{sarnak06}
\begin{equation}\label{eq:future-riesz}
\begin{cases}
\mathrm{BCC}, & 0<s<3/4,\\
\mathrm{BCC}\ \text{and}\ \mathrm{FCC}, & s=3/4,\\
\mathrm{FCC}, & 3/4<s<3/2,
\end{cases}
\end{equation}
with uniqueness up to isometries away from the transition. The
crossing at $s=3/4$ follows from duality; global optimality there
remains conjectural. Sarnak and Str\"ombergsson established the FCC
comparison for the height of flat three-tori at the pole $s=3/2$
\cite{sarnak06}, and Luo and Wei proved FCC optimality for $s>3/2$
\cite{luo26}. Thus the interval $0<s<3/2$, apart from the dual pair
\(s=1/2\) and \(s=1\) settled here, remains a natural target.

\paragraph{More general periodic crystals.}
Another extension of Theorem~\ref{thm:main} is to periodic sets
$\bigcup_{j=1}^{m}(L+a_j)$ with $m\geq2$ identical particles per
period cell.  This includes hexagonal close packing (HCP) lattice,
which is not a Bravais lattice but has a two-point basis over a
hexagonal lattice. Already for $m=2$, an unrestricted configuration
has five lattice-shape parameters and three relative-position
parameters, rather than the five parameters considered here. Numerical
Madelung energies favor BCC over HCP \cite{kozhberov21}. Proving the
corresponding comparison uniformly for two-point, and eventually
larger finite bases would be an important intermediate step towards
relaxing the overall periodicity assumption imposed by lattice $L$.
Ultimately, the question is whether BCC minimizes the renormalized
Coulomb energy among configurations without any periodicity
assumption. This is a substantially harder three-dimensional
crystallization problem; see \cite{blanc15,rougerie16,lewin22}.

\paragraph{Ionic crystals.}
Yet another natural generalization would be to consider lattices with
several {\em distinct} charges of the same sign per lattice period
that model, for example, mixtures of different atomic nuclei in a
uniform neutralizing electron background at ultra-high densities of
nuclear matter. Here BCC lattices with the larger charge in the center
of each elementary cubic cell are natural candidates as minimizers for
two charge types when the charge contrast is not too high. At the same
time, at strong contrast one would expect the high charges themselves
to form a BCC lattice. Notice that from the point of view of the
macroscopic ground states one would also need to consider {\em phase
  separated} configurations consisting of each charge type organizing
into its own BCC lattice. A further question concerns charges of {\em
  different signs}, whose classical application is the crystalline
structure of ionic crystals. Here, however, the original model would
fail to produce energy minimizers, as the energy would go to minus
infinity when the opposite charges approach one another. Therefore,
one needs to introduce an additional short-range repulsion between all
the charges, such as, for example, a repulsive Riesz potential with
high exponent (e.g. the standard soft core potential with $s = 6$ in
our exponent convention). The volume of the unit cell now also needs
to be determined, considering the energy minimality against lattice
dilations. This problem goes back to the classical works
\cite{madelung18,born21} and would provide a rich playground for
further applications of our methodology (for some recent studies, see
\cite{betermin18,betermin21}).

\paragraph{Acknowledgements} The authors are members of INdAM/GNAMPA
and acknowledge the MUR Excellence Department Project awarded to the
Department of Mathematics, University of Pisa, CUP
I57G22000700001. C. B. M. would like to thank L. B\'etermin for
valuable discussions at an early stage of this project.

\appendix

\renewcommand{\thetheorem}{\thesection\arabic{theorem}}
\setcounter{theorem}{0}

\section{The zero screening limit}\label{app:screened}

\begin{lemma}\label{lem:screened}
  Let $L$ be a lattice of covolume one and let $v$ be the zero-mean
  $L$-periodic solution of \eqref{eq:vOmL}. Then the limit in
  \eqref{eq:EL} exists, equals
  $\lim_{x\to0}\bigl(v(x)-|x|^{-1}\bigr)$, and \eqref{eq:KLEL}
  holds. Moreover
  \begin{align}
    \label{eq:kappato0}
    \sum_{p\in L\setminus\{0\}}\frac{e^{-\kappa|p|}}{|p|}
    -\frac{4\pi}{\kappa^2} - \kappa 
    \ =\ \lim_{x\to0}\left( v(x)-\frac1{|x|} \right)  + O(\kappa^2)
    \qquad \text{as} \ \kappa\to0^+.
  \end{align}
\end{lemma}

\begin{proof}
  Since $e^{-\kappa r}/r$ is the Green's function of
  $-\Delta+\kappa^2$ with the $4\pi$ normalization, the function
  \begin{align}\label{eq:scr-vk}
    v_\kappa(x) := \sum_{p\in
                  L}\frac{e^{-\kappa|x-p|}}{|x-p|}-\frac{4\pi}{\kappa^2}
  \end{align}
  solves
  \begin{align}
    \label{eq:screened}
    (-\Delta+\kappa^2)v_\kappa = 4\pi\Bigl(\sum_{p\in L}\delta_p-1\Bigr).
  \end{align}
  The series in \eqref{eq:scr-vk} converges absolutely and locally
  uniformly off $L$ for every $\kappa>0$.  Because $\Omega_L$ tiles
  $\mathbb R^3$ and has unit volume,
  \begin{align}\label{eq:scr-mean}
    \int_{\Omega_L}\sum_{p\in L}\frac{e^{-\kappa|x-p|}}{|x-p|}\,dx
    &=\int_{\mathbb R^3}\frac{e^{-\kappa|y|}}{|y|}\,dy
    =4\pi\int_0^\infty re^{-\kappa r}\,dr=\frac{4\pi}{\kappa^2},
  \end{align}
  so $v_\kappa$ has mean zero over $\Omega_L$, the same normalization
  as $v$.

  For an $L$-periodic locally integrable function
  $f: \mathbb R^3 \to \mathbb R$, we now introduce the Fourier
  representation \cite{conway}
  \begin{align}
    \label{eq:fourier}
    f(x) = \sum_{q \in L^*} e^{2 \pi i q \cdot x} \widehat f(q), \qquad
    \widehat f(q) = \int_{\Omega_L} e^{-2 \pi i q \cdot x} f(x) \, dx,
  \end{align}
  where $L^*$ is the dual lattice to $L$.  With this convention, for
  every nonzero $q \in L^{*}$ we have
  \begin{align}\label{eq:scr-fourier}
    \widehat v(q) &=\frac{1}{\pi|q|^2}, &
    \widehat v_\kappa(q) &=\frac{4\pi}{4\pi^2|q|^2+\kappa^2},
  \end{align}
  while $\widehat v(0)=\widehat v_\kappa(0)=0$. Hence
  \begin{align}\label{eq:scr-diff}
    \bigl|\widehat v(q)-\widehat v_\kappa(q)\bigr|
    &=\frac{\kappa^2}{\pi|q|^2\bigl(4\pi^2|q|^2+\kappa^2\bigr)}
    \ \le\ \frac{\kappa^2}{4\pi^3|q|^4} .
  \end{align}
  Since
  $\displaystyle \sum_{q\in L^* \backslash \{0\}}|q|^{-4}<\infty$ in
  dimension three, the Fourier series of $v-v_\kappa$ converges
  absolutely; therefore, $v-v_\kappa$ is continuous, may be evaluated
  at the origin, and
  \begin{align}
    \|v-v_\kappa\|_\infty\ \le\ \frac{\kappa^2}{4\pi^3}
    \sum_{q\in L^*\setminus\{0\}}\frac1{|q|^4}\ =\ O(\kappa^2).    
  \end{align}
  In particular the singular parts of $v$ and $v_\kappa$ at the origin
  agree and cancel.

  Finally, isolating the term $p=0$ in \eqref{eq:scr-vk} and using
  \begin{align}\label{eq:scr-self}
    \lim_{x\to0}\Bigl(\frac{e^{-\kappa|x|}}{|x|}-\frac1{|x|}\Bigr) &= -\kappa ,
  \end{align}
  we obtain
  \begin{align}
    \lim_{x\to0}\Bigl(v_\kappa(x)-\frac1{|x|}\Bigr) =\sum_{p\in
    L\setminus\{0\}}\frac{e^{-\kappa|p|}}{|p|}-\frac{4\pi}{\kappa^2}-\kappa
    .
  \end{align}
  Letting $\kappa\to0^+$ and using $\|v-v_\kappa\|_\infty=O(\kappa^2)$
  shows that the limit in \eqref{eq:EL} exists and equals
  $\lim_{x\to0}\bigl(v(x)-|x|^{-1}\bigr)$. Comparing with
  \eqref{eq:Madelung} gives \eqref{eq:KLEL}.
\end{proof}

\bibliographystyle{plain}

\bibliography{../../nonlin,../../stat,../../mura}

\end{document}